\documentclass[prb,twocolumn,showpacs,superscriptaddress]{revtex4}

\usepackage{graphicx}
\usepackage{dcolumn}
\usepackage{amsmath}
\usepackage{amsfonts}
\usepackage{amssymb}
\usepackage{bm}
\usepackage{epsfig}

\begin{document}


\title{Neutron scattering from a coordination polymer quantum paramagnet}

\author{T. Hong}
\affiliation{Department of Physics and Astronomy, The Johns Hopkins
University, Baltimore, Maryland 21218, USA}
\author{M. Kenzelmann}
\affiliation{Department of Physics and Astronomy, The Johns Hopkins
University, Baltimore, Maryland 21218, USA} \affiliation{National
Institute of Standards and Technology, Gaithersburg, Maryland 20899,
USA}
\author{M. M. Turnbull}
\affiliation{Carlson School of Chemistry and Department of
Physics, Clark University, Worcester, Massachusetts 01610, USA}
\author{C. P. Landee}
\affiliation{Carlson School of Chemistry and Department of
Physics, Clark University, Worcester, Massachusetts 01610, USA}
\author{B. D. Lewis}
\affiliation{Carlson School of Chemistry and Department of Physics,
Clark University, Worcester, Massachusetts 01610, USA}
\author{K. P. Schmidt}
\affiliation{Institute of Theoretical Physics, \'{E}cole
Polytechnique F\'{e}d\'{e}rale de Lausanne, CH 1015 Lausanne,
Switzerland}
\author{G. S. Uhrig}
\affiliation{Theoretische Physik, FR 7.1, Geb. 38, Universit\"{a}t
des Saarlandes, 66123 Saarbr\"{u}cken, Germany}
\author{Y. Qiu}
\affiliation{National Institute of Standards and Technology,
Gaithersburg, Maryland 20899, USA} \affiliation{Department of
Materials and Engineering, University of Maryland, College Park,
Maryland 20742, USA}
\author{C. Broholm}
\affiliation{Department of Physics and Astronomy, The Johns Hopkins
University, Baltimore, Maryland 21218, USA} \affiliation{National
Institute of Standards and Technology, Gaithersburg, Maryland 20899,
USA}
\author{D. H. Reich}
\affiliation{Department of Physics and Astronomy, The Johns
Hopkins University, Baltimore, Maryland 21218, USA}


\date{\today}

\begin{abstract}
Inelastic neutron scattering measurements are reported for a powder
sample of the spin-1/2 quantum paramagnet $\rm Cu(Quinoxaline)Br_2$.
Magnetic neutron scattering is identified above an energy gap of 1.9
meV. Analysis of the sharp spectral maximum at the onset indicates
that the material is magnetically quasi-one-dimensional.
Consideration of the wave vector dependence of the scattering and
polymeric structure further identifies the material as a two-legged
spin-1/2 ladder. Detailed comparison of the data to various models
of magnetism in this material based on the single mode approximation
and the continuous unitary transformation are presented. The latter
theory provides an excellent account of the data with leg exchange
$J_{\parallel}=2.0$\ meV and rung exchange $J_{\perp}=3.3$\ meV.
\end{abstract}

\pacs{75.10.Jm, 75.40.Gb, 75.50.Ee}
\maketitle

\section{INTRODUCTION}
Progress towards understanding the cooperative quantum physics of
one dimensional systems is frequently gated by the availability of
suitable model systems for experiments. While oxides offer the
possibility of carrier doping, coordination polymer magnets have
energy scales that are well suited for high magnetic field
experiments that traverse phase boundaries. Such experiments have
been important for elucidating the fermionic nature of low energy
excitations in the uniform spin-1/2 chain \cite{Stone03:91} and
their remarkable transition to a gapped solitonic spectrum upon
application of an effective staggered field.\cite{Kenze04:93}

\begin{figure}[t]
\epsfig{width=3.5cm,file=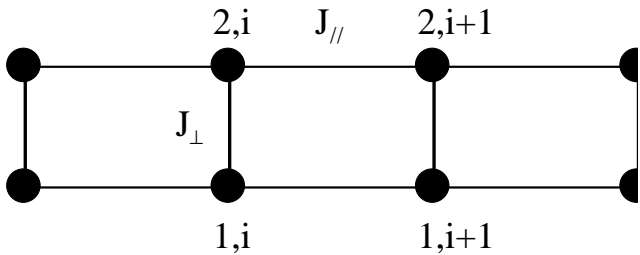,bbllx=115,bblly=105,bburx=285,bbury=530,angle=-90,clip=}
\caption{Schematic of a two-leg spin ladder system, indicating the
rung interaction $J_{\perp}$ and the rail interaction
$J_{\parallel}$. The indexing of spin sites used in
Eq.~(\ref{ladeqn}) is also shown.} \label{ladder}
\end{figure}

A natural next step in the experimental exploration of one
dimensional spin-1/2 systems is to examine magnetic excitations as a
function of field and temperature in two legged spin-1/2 ladders.
The Hamiltonian of this system has the form

\begin{eqnarray} \label{ladeqn}
H &=& \sum_{i} J_{\perp} {\bf S}_{1,i}\cdot {\bf S}_{2,i}\\
\nonumber &+&J_{\parallel}({\bf S}_{1,i}\cdot {\bf S}_{1,i+1}+{\bf
S}_{2,i}\cdot {\bf S}_{2,i+1}),
\end{eqnarray}
where \emph{i} indexes rungs and 1, 2 each leg as illustrated in
Fig. $\ref{ladder}$. The Lieb-Schultz-Mattis theorem\cite{Lieb61:16}
allows for a finite gap in the excitation spectrum and indeed
analytical,\cite{Barnes93:47,uhrig96b,sushk98,damle98,jurecka00,trebs00,knett01}
numerical,\cite{Barnes93:47} and experimental\cite{Azuma94:73} work
has confirmed that a spin gap does exist. A key experiment is to
close this spin gap with an external field and explore the predicted
quantum critical high field phase.\cite{Muller81:24} A rich zero
field excitation spectrum that should feature both bound states and
continua also remains to be examined
experimentally.\cite{knett01,Barnes03:67}

Unfortunately as of now no experimental model system has been
identified to enable neutron scattering experiment in the high field
phase. $\rm (La,Sr,Ca)_{14}Cu_{24}O_{41}$\cite{Carter96:77} and $\rm
SrCu_2O_3$\cite{Azuma94:73} do appear to be spin ladders but the
high energy scales for these systems render them unsuitable for the
proposed experiments. The energy scales for $\rm
Cu_2(1,4-diazacycloheptane)_2Cl_4$ (CuHpCl)
\cite{Chiari90:29,Hammar96:79,Chaboussant97:55,Chaboussant97:79,Hammar97:57}
and $\rm (VO)_2P_2O_7$\cite{Johnston87:35,Eccleston94:73} are
appropriate and they were previously identified as spin-1/2 ladders
on the basis of specific heat and magnetization measurements.
However, further investigations with neutron scattering revealed
that the exchange interactions in these materials do not correspond
to spin ladders despite spin ladder structural motifs. In the case
of CuHpCl there were early indications from magnetization
measurements that the material is not one dimensional and subsequent
neutron scattering experiments showed that the group velocity for
magnetic excitation has no substantial anisotropy.\cite{Stone02:65}
In addition, the molecular bond, which was thought to be the rung of
the ladder, assumes a frustrated configuration in the ground state.
Hence CuHpCl is now thought to have a complex frustration induced
singlet ground state.\cite{Stone02:65} $\rm (VO)_2P_2O_7$ on the
other hand is quasi-one-dimensional but rather than being a spin
ladder, neutron scattering experiments showed that it is an
alternating spin chain with the chain direction perpendicular to the
putative ladder
direction.\cite{Garrett97:55,Tennant97:78,Garrett97:79}

These false starts illustrate that bulk measurements, while
sensitive to the magnetic density of states, can not provide
definite conclusions regarding the geometry of interacting spin
system with a singlet ground state. With the goal of identifying
spin ladder systems for high field experiments, we have initiated a
project to synthesize potential coordination polymer spin ladders
and subject them to inelastic neutron scattering experiments for
verification. $\rm Cu(Quinoxaline)Br_2$ is a first out of several
such candidate materials.\cite{Watson} Susceptibility measurements
on this material \cite{Landee03:22} are well described by a
spin-ladder model with antiferromagnetic exchange constants
$J_{\parallel}=2.61$ meV and $J_{\perp}=3.02$ meV. Magnetization
measurements \cite{Landee03:22} show a phase transition from a
non-magnetic singlet ground state to a magnetized state near a
critical field $H_c=14$ T, indicating the presence of an energy gap
$\Delta=g\mu_BH_c\approx 1.7$ meV. Furthermore the square root
singular onset of magnetization above the critical field suggests
that the system is magnetically one-dimensional.

In the present paper, we report inelastic neutron scattering
experiment on a deuterated powder sample of $\rm
Cu(Quinoxaline)Br_2$ to determine whether this material is indeed a
spin ladder and to establish the strength of the relevant exchange
interactions. The comparison of wave vector and energy dependent
magnetic neutron scattering data to comprehensive model calculations
carried out as a function of rung and leg exchange constants
strongly indicate that $\rm Cu(Quinoxaline)Br_2$ is a coordination
polymer spin ladder, with weak inter-ladder interactions. Final
verification will require inelastic scattering experiments from
single crystals.

\begin{figure}[t]
\centering
\epsfig{width=8cm,file=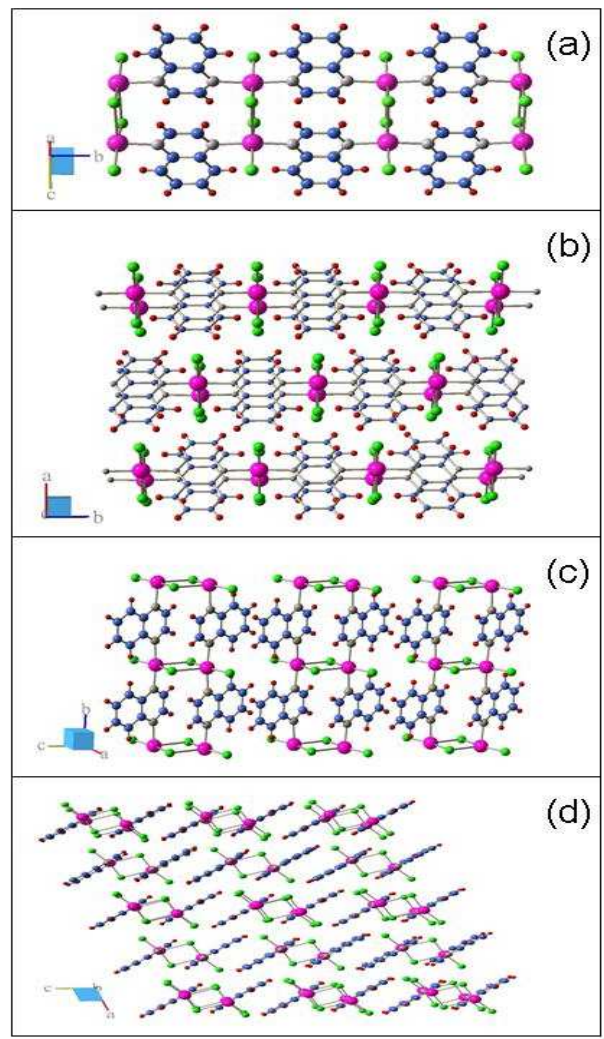,angle=0,bbllx=90,bblly=55,bburx=500,bbury=765,clip=}
\caption{ Structure of $\rm Cu(Quinoxaline)Br_2$ projected along
four different directions to illustrate potential interacting spin
models. (a) The molecular ladder structure of $\rm
Cu(Quinoxaline)Br_2$\cite{Landee03:22}, which the ladder extends
along the monoclinic {\bf b} axis. Interactions along the
$(\bf{a}\pm\bf{b})/2$ and $\bf{c}$ directions could lead to two
dimensional dimer models in the $\bf{a}$-$\bf{b}$ or
$\bf{b}$-$\bf{c}$ planes (frame (b) and (c)) or a three dimensional
dimer model (frame (d)). Color coding is as follows: Pink:Cu,
Green:Br, Blue:C, Red: H, and Grey:N.} \label{quino}
\end{figure}

The crystal structure of $\rm Cu(Quinoxaline)Br_2$ ($\rm
Cu(C_8H_6N_2)Br_2$) is monoclinic with space group C2/m and lattice
constants $\rm a=13.1745(15)$ $\rm \AA$, $\rm b=6.9293(8)$ $\rm
\AA$, $\rm c=10.3564(12)$ $\rm \AA$, and
$\beta=107.699(2)^\circ$.\cite{Landee03:22} Quinoxaline ($\rm
C_8H_6N_2$) has a tendency towards polymeric structures where
quinoxaline forms a bridge between metal atoms.\cite{Lumme87:43} In
$\rm Cu(Quinoxaline)Br_2$, $\rm Cu_2Br_4$ dimers are linked to
adjacent dimers by bridging quinoxaline molecules along the
monoclinic {\bf b} axis as shown in Fig.~\ref{quino}(a). The
inter-dimer spacing along the {\bf b}-axis is 6.929 $\rm\AA$.
Hydrogen bond mediated interactions may exist between molecular
units displaced by $\bf{d}_1=(\bf{a}\pm\bf{b})/2$ and
$\bf{d}_2=\bf{c}$ (See Fig.~\ref{quino}). If there were sufficiently
strong, the material could be a two-dimensional bi-layer system as
in $\rm BaCuSi_2O_6$,\cite{Sasago97:55} an alternating spin chain as
in $\rm Cu(NO_3)_2\cdot 2.5 D_2O$,\cite{Xu00:84} or a three
dimensional system of interacting spin pairs as in
TlCuCl$_3$\cite{Cavadini01:63} and
KCuCl$_3$.\cite{Cavadini99:7,Cavadini00:12}

\section{EXPERIMENTAL TECHNIQUES}

\subsection{Sample and Neutron Instrumentation}

The neutron scattering sample consisted of 27 grams of deuterated
powder in an annular aluminum can with inner diameter 2.42 cm, outer
diameter 2.92 cm, and height 10 cm. The powder was obtained through
precipitation of a 1:1 molar solution of anhydrous $\rm CuBr_2$ and
deuterated d-4 quinoxaline dissolved in a small amount of 95\%
ethanol. Prompt gamma neutron activation analysis and NMR
measurements showed that 58\% of the hydrogen sites in our sample
were occupied by deuterium. This is less than the 2/3 deuteration
level of the d-4 quinoxaline starting material indicating some
hydrogen/deuterium exchange with the solvent.

Inelastic neutron scattering measurements were performed using the
disk chopper time-of-flight spectrometer (DCS)\cite{Copley03} at the
National Institute of Standards and Technology (NIST) Center for
Neutron Research in Gaithersburg, Maryland. A disk chopper system
was used to select a 167 Hz pulsed neutron beam with an energy of
4.87 meV and a pulse width of 79 $\mu$s from the NCNR cold neutron
source. The 0.65 steradian detection system probed inelastic
scattering with energy transfer $-4.17~\rm
meV~$$\leq\hbar\omega\leq3.84$ meV and momentum transfer 0.13 $\rm
\AA^{-1}\leq Q\leq$ 3.40 $\rm \AA^{-1}$. The full width at half
maximum (FWHM) elastic energy resolution was
$\delta\hbar\omega\simeq0.18$ meV.

\subsection{Neutron Scattering Data Analysis}
\label{nda} The remnant hydrogen in the sample gives rise to a
scattering cross section of 201 barn per formula unit compared to a
total magnetic scattering cross section from Cu$^{2+}$  of 0.146
barn per formula unit. There are several distinct challenges
associated with the strong incoherent nuclear scattering. (1) The
strong elastic scattering cross section reveals tails of the energy
resolution function that can dwarf the magnetic signal well beyond
the FWHM of the resolution function. (2) Phonon scattering and
multiple scattering processes involving phonons produce a strong
background even at low temperatures that must be determined
accurately in order to isolate the magnetic scattering. (3) Both
magnetic and phonon scattering are frequently preceded or followed
by incoherent elastic nuclear scattering. Because incoherent
scattering is approximately wave vector independent this leads to a
$Q$-averaged contribution of both types of scattering at any wave
vector. We have developed accurate techniques for dealing with each
of these issues and because they are essential for this research
they are described in the following sections.

\subsubsection{Temperature independent background}
The elastic background from the tails of the resolution function can
be determined by utilizing the fact that inelastic scattering obeys
the principle of detailed balance whereas elastic scattering at
sufficiently low temperatures, to a good approximation, can be
approximated as being temperature independent. Under these
circumstances the raw measured count rate, $I_r(\omega,T)$, for any
specific value of wave vector transfer satisfies:
\begin{eqnarray}
I_r(\omega,T)&=&B(\omega)+I(\omega,T).\label{detb1}\\
I_r(-|\omega|,T)&=&B(-|\omega|)+I(|\omega|,T)\exp(-\hbar|\omega|/kT).\label{detb2}
\end{eqnarray}
Here $B(\omega)$ is the temperature independent
background and $I(\omega,T)$ represents all inelastic scattering
processes that satisfy detailed balance at the temperature $T$. An
underlying assumption here is that resolution effects can be
neglected for the inelastic scattering. By measuring data at two
different temperatures $T_1$=1.4 K and $T_2$=60 K and using
equations (\ref{detb1}) and (\ref{detb2}) it is possible to
extract values for $B(|\omega|)$, $B(-|\omega|)$, $I(\omega,T_1)$
and $I(\omega,T_2)$ from the corresponding four equations.

\subsubsection{Subtracting phonon scattering}
The above procedure isolates inelastic scattering from temperature
independent elastic scattering and detector dark current. However,
this scattering intensity still has contributions from magnetic as
well as phonon scattering. To remove the low temperature
contribution from phonon scattering, $I_p(\omega,T_1)$ ,  and
hence isolate low temperature magnetic scattering,
$I_m(\omega,T_1)$, we use the fact that phonon scattering in
hydrogenous systems dwarfs magnetic scattering for
$k_BT>>\hbar\omega$. Hence we make the following approximation for
$k_BT_2>\hbar\omega$:
\begin{equation}
I(\omega,T_2)\equiv I_{p}(\omega,T_2)+I_{m}(\omega,T_2)\approx
I_{p}(\omega,T_2).
\end{equation}

We further assume that one phonon scattering events dominate over
multi-phonon events and we neglect the temperature dependence of the
Debye Waller factor such that\cite{Lovesey}
\begin{equation}
I_{p}(\omega,T_1)\approx
\frac{1-\exp(-\beta_2\hbar\omega)}{1-\exp(-\beta_1\hbar\omega)}I(\omega,T_2)\label{tact}.
\end{equation}

The final expression for the magnetic inelastic scattering as deduced from the measurements at $T_1=1.5$~K and $T_2=60$~K is therefore
\begin{equation}
I_{m}(\omega,T_1)\approx
I(\omega,T_1)-\frac{1-\exp(-\beta_2\hbar\omega)}{1-\exp(-\beta_1\hbar\omega)}I(\omega,T_2).
\end{equation}

\subsubsection{Multiple Scattering}

Multiple scattering is difficult to avoid when probing weak magnetic
scattering in the presence of strong incoherent scattering. Here we
describe how to account for the dominant double scattering process
that involves incoherent elastic scattering preceded or followed by
a weaker inelastic process. For simplicity we assume that the
incoherent elastic scattering is isotropic and independent of
neutron energy and we neglect anisotropy introduced by the sample
geometry.\cite{aniso} Under those circumstances the incoherent
scattering event in a double scattering process effectively
randomizes the direction of scattering for the preceding or
subsequent inelastic scattering:
\begin{eqnarray}
I^{\prime}(Q,\omega)&=&{\cal T} I(Q,\omega)\nonumber\\
&&+(1-{\cal
T})\int_{|k_i-k_f|}^{k_i+k_f}I(Q^{\prime},\omega)\frac{Q^{\prime}dQ^{\prime}}{2k_ik_f}.
\label{multip}
\end{eqnarray}
Here $k_i$ and $k_f$ are the incident and scattered wave vectors
respectively. The integral implements the average over scattering
directions for the inelastic process. $\cal T$ is an ``effective"
sample transmission that can be determined through analysis of
phonon scattering. Approximating the intensity of single event
phonon scattering for a thin sample as $I_p(Q,\omega )=f(\omega)Q^2$
we find that
\begin{equation}
I_p^{\prime}(Q,\omega)=f(\omega)(k_i^2+k_f^2)({\cal
T}\frac{Q^2}{k_i^2+k_f^2}+(1-{\cal T})). \label{phonon}
\end{equation}

\begin{figure}
\includegraphics[width=9cm,bbllx=95,bblly=375,bburx=545,
  bbury=685,angle=0,clip=]{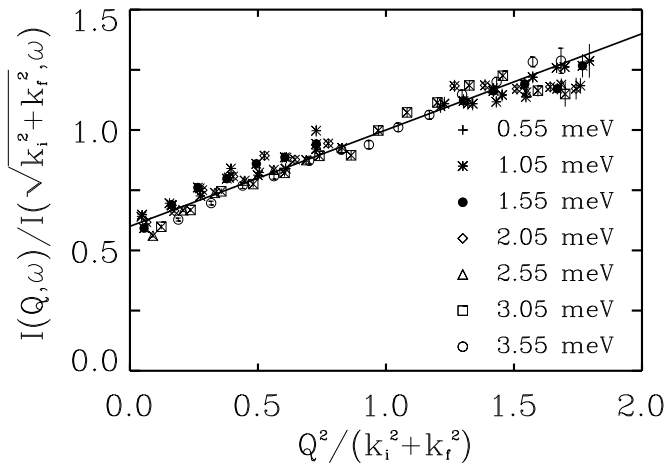}
\caption{Scaled wave vector dependence of inelastic neutron
scattering from $\rm Cu(Quinoxaline)Br_2$ at $T$=60 K and for
various values of energy transfer. The scaling behavior is
consistent with double scattering by phonons and incoherent elastic
nuclear scattering as described by Eq.~(\ref{multip})
and~(\ref{phonon}).} \label{mul}
\end{figure}

Fig.~\ref{mul} shows phonon scattering intensity at $T=60$ K as a
function of $Q^2/(k_i^2+k_f^2)$. Data for several values of energy
transfer, $\hbar\omega$, scale in agreement with Eq.~(\ref{phonon})
and yield an effective value of ${\cal T}=0.40$. For comparison the
average transmission through a spherical version of our sample as
calculated considering only incoherent scattering is 0.30. In the
following we shall use Eq.~\ref{multip} with ${\cal T}=0.40$ when
comparing the measured inelastic magnetic scattering, $I_m(\omega,
T)$, to theoretical models of spin dynamics in $\rm
Cu(Quinoxaline)Br_2$.

\subsubsection{Normalized magnetic scattering}
Anisotropic self shielding effects associated with the annular
sample geometry were taken into account using a numerical
integration technique.\cite{ITC04:C} Absolute normalization of the
data was subsequently achieved using elastic incoherent scattering
from the sample, duly considering the hydrogen/deuterium ratio
determined through neutron activation and NMR
analysis.\cite{Hammar97:57} The technique is estimated to be
accurate to within 20\%. The normalized magnetic scattering
intensity, $\tilde{I}_m$, is related to the resolution smeared
dynamic spin correlation function as follows:
\begin{eqnarray}
\tilde{I}_m(Q,\omega ) &=& 2 \int dQ^{\prime}  \hbar d\omega
^{\prime}
{\cal R}_{Q\omega }(Q-Q^{\prime},\omega-\omega^{\prime})\nonumber\\
&&\times |\frac{g}{2}F({\bf{Q}}^{\prime})|^2{\cal
\widetilde{S}}(Q^{\prime},\omega^{\prime} ). \label{inorm}
\end{eqnarray}

Here ${\cal R}_{Q\omega }(Q-Q^{\prime},\omega-\omega^{\prime})$ is a
unity normalized resolution function that is  peaked on the scale of
the FWHM resolution for $Q\approx Q^{\prime}$ and
$\hbar\omega\approx \hbar\omega^{\prime}$, $g\simeq 2.12$ is the
$\rm Land$\'e{} g-factor for $\rm {Cu^{2+}}$ in $\rm
Cu(Quinoxaline)Br_2$.\cite{Landee03:22} $\textit{F}({\bf{Q}})$ is
the anisotropic magnetic form factor appropriate for a hole in the
$3d_{x^2-y^2}$ orbital.\cite{shamoto93:48} Based on bond distances
and their coplanarity it appears that the two short $\rm Cu$-$\rm
Br$ bonds and the two bonds to the neighboring quinoxaline molecules
define the $\hat{\bf x}$-$\hat{\bf y}$ plane for the $3d_{x^2-y^2}$
copper orbitals. We used this assumption to model the effects of the
anisotropic magnetic form factor. The spherical average of the
dynamic spin correlation function is
\begin{equation}
\widetilde{{\cal S}}(Q,\omega)=\int \frac{d\Omega_{\hat{Q}}}{4\pi}
\frac{1}{2} \sum_{\alpha\beta}
(\delta_{\alpha\beta}-\hat{Q}_\alpha\hat{Q}_\beta) {\cal
S}^{\alpha\beta}({\bf Q},\omega), \label{powderavg}
\end{equation}
and the dynamic spin correlation function proper is
\begin{eqnarray}
\label{eq_sqw}
{\cal S}^{\alpha\beta}({\bf Q},\omega)&=&\frac{1}{2\pi\hbar}
\int dte^{i\omega t}\frac{1}{N} \sum_{{\bf R}{\bf R}^{'}}\nonumber\\
&&<S^\alpha_{{\bf R}}(t)S^\beta_{{\bf R}^{'}}(0)>
e^{-i{\bf Q} \cdot ({\bf R}-{\bf R}^{'})} .
\end{eqnarray}

Unless specifically stated, we use the notation defined in S. M.
Lovesey's book on thermal neutron scattering.\cite{Lovesey}

\section{Experimental Results}
\begin{figure}
\includegraphics[width=8.5cm,bbllx=99,bblly=122,bburx=457,
  bbury=737,angle=0,clip=]{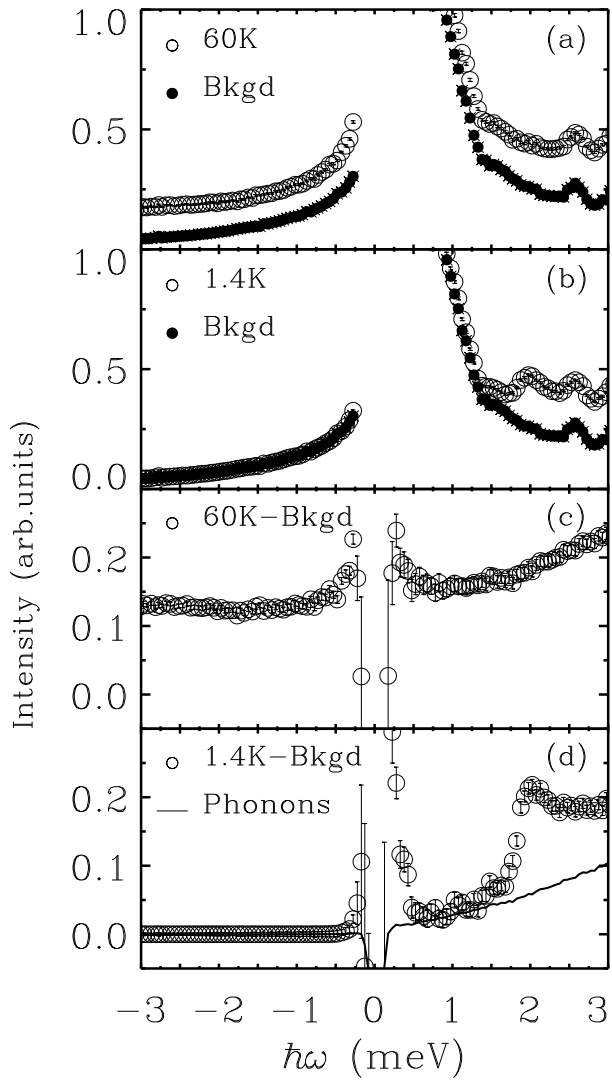}
\leavevmode \caption{The energy dependence of Q-integrated neutron
scattering from $\rm Cu(Quinoxaline)Br_2$ at (a) T=60 K and (b)
T=1.4 K before detailed balance correction. In both frames solid
symbols show the temperature independent background determined as
described in section $\amalg$.B,1. Background subtracted data at (c)
T=60 K and (d) T=1.4 K data. The solid line is phonon scattering
contribution at T=1.4 K as determined by Eq.~(\ref{tact}).}
\label{rawdata}
\end{figure}

Fig.~\ref{rawdata} (a-b) shows raw Q-integrated data at T=60 K and
T=1.4 K. The range covered is 0.5 $\rm \AA^{-1}\leq Q\leq$ 1.0 $\rm
\AA^{-1}$. The solid symbols in Fig.~\ref{rawdata}(b) show
$B(\omega)$ as determined by Eq.~(\ref{detb1}) and
Eq.~(\ref{detb2}). It is clear from this analysis that the strong
incoherent elastic scattering from hydrogen produces intense tails
of scattering intensity well beyond the 0.18 meV FWHM of the
resolution function. The increase of the count rate beyond 3 meV
(see Fig.~\ref{rawdata}(a)) is attributable to the same effect
through time of flight frame overlap.

A small peak is visible in the raw low temperature data
(Fig.~\ref{rawdata}(b)) at approximately 2 meV. Following background
subtraction, Fig.~\ref{rawdata}(d) clearly shows that this peak
marks the onset of a continuum of scattering in the wave vector
integrated spectrum. Fig.~\ref{rawdata}(c) shows that this continuum
is not present at T=60 K where as expected, thermally activated
phonon scattering dominates. The solid line in Fig.~\ref{rawdata}(d)
shows the projected contribution from phonon scattering at 1.4 K as
derived from the data in Fig.~\ref{rawdata}(c) using
Eq.~(\ref{tact}). While incoherent inelastic phonon scattering
produces a background contribution that increases with
$\hbar\omega$, it is apparent that the bounded continuum above a
$\sim2$ meV spectral gap is associated with inelastic magnetic
scattering. This spin-gap is consistent with the activation energy
derived from bulk measurements.\cite{Landee03:22}

\begin{figure}
\includegraphics[width=9cm,bbllx=100,bblly=115,bburx=585,
  bbury=685,angle=0,clip=]{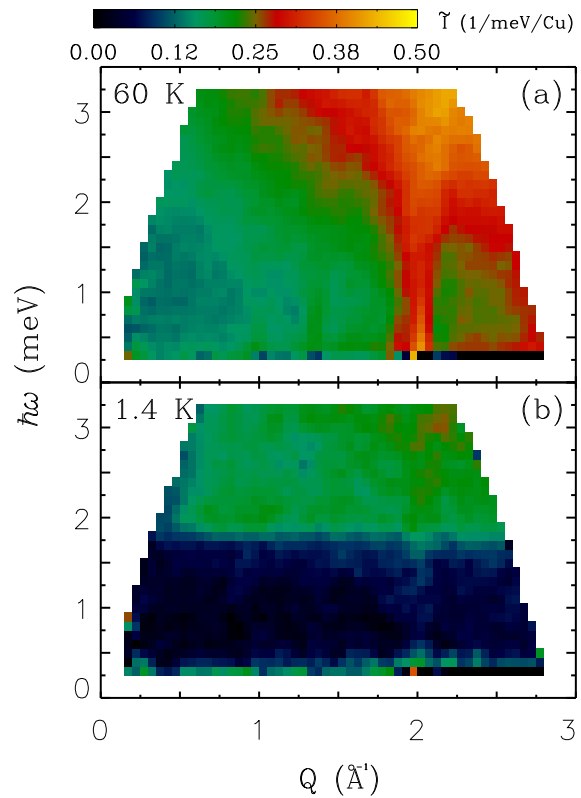}
\caption{Normalized powder inelastic neutron scattering intensity
         for $\rm Cu(Quinoxaline)Br_2$
         at (a) T=60 K and (b) T=1.4 K after detailed balance correction as described in section $\amalg$.B,1.}
\label{contour1}
\end{figure}

Having verified that magnetic scattering can be isolated from
incoherent nuclear scattering through the techniques described in
section \ref{nda}, the analysis can now be replicated as a function
of wave vector transfer. The results are summarized in
Figs.~\ref{contour1} and~\ref{contour2}, where the data have
furthermore been normalized to report values for
$\widetilde{I}(Q,\omega)$  as defined by Eq.~(\ref{inorm}).

Fig.~\ref{contour1} shows the normalized scattering intensity
$\widetilde{I}(Q,\omega)$ for $\rm Cu(Quinoxaline)Br_{2}$ at T=1.4 K
and T=60 K after subtracting the temperature independent background
determined through Eq.~({\ref{detb1}) and Eq.~(\ref{detb2}). At T=60
K the scattering increases with wave vector transfer as expected for
phonon scattering. Further analysis of the $Q-$dependence of these
data was provided in section~\ref{nda}.3. At low temperatures there
is a clear onset of scattering for $\hbar\omega>2$~meV and this is
consistent with magnetic neutron scattering from a quantum
paramagnet. However, a growth in the scattering intensity towards
higher $Q$ confirms the conclusions from the previous spectral
analysis that even the low $T$ scattering contains significant
contributions from phonon scattering.

Fig.~\ref{contour2}(a) shows the phonon contribution to low
temperature scattering derived from the T=60 K data based on
Eq.~(\ref{tact}). Subtracting these data from the data in
Fig.~\ref{contour1}(b) concludes the process of isolating the $Q$
and $\hbar\omega$ dependent magnetic scattering. From these data,
which are displayed in Fig.~\ref{contour2}(b), it is apparent that
there is a sharp onset of magnetic scattering for
$\hbar\omega\simeq1.9$ meV with a finite-$Q$ maximum. In the
following we shall show that these are features of a
quasi-one-dimensional antiferromagnet with a singlet ground state
and that the data furthermore are consistent with a spin ladder
extending along the {\textbf b}-axis.

\begin{figure}
\includegraphics[width=9cm,bbllx=100,bblly=115,bburx=585,
  bbury=685,angle=0,clip=]{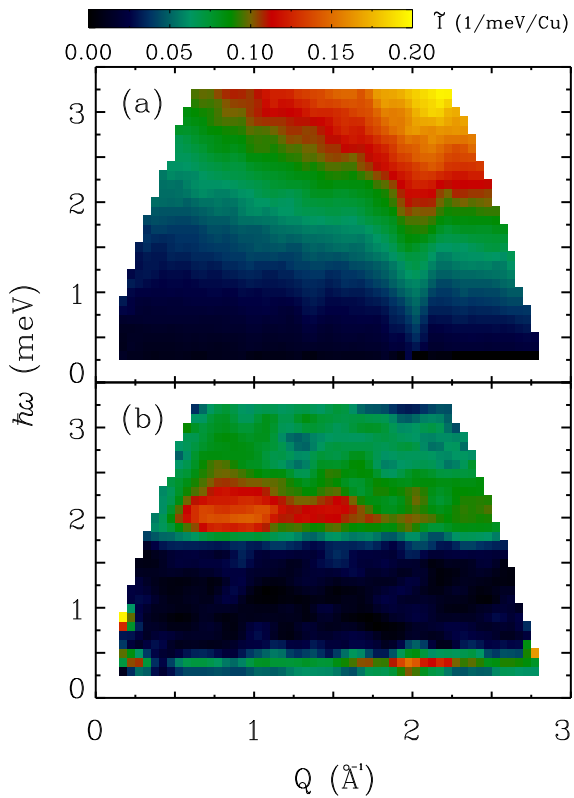}
\caption{Normalized inelastic scattering intensity for $\rm
Cu(Quinoxaline)Br_{2}$
         (a) Phonon scattering intensity at T=1.4 K.
         (b) Magnetic scattering intensity at T=1.4 K after subtracting
          the phonon contribution as described in section $\amalg$.B,1.}
\label{contour2}
\end{figure}

\section{ANALYSIS and DISCUSSION}

\subsection{Single Mode Approximation}
For most but not all\cite{Shastry81:47} quantum magnets, the
dominant spectral weight resides in a resonant mode. When this
excitation is between a singlet ground state and an excited triplet
without magnetic long range order, it is called a
triplon.\cite{schmi03c} We prefer to reserve the common term magnon
for systems with magnetic long range order. The "single mode
approximation" (SMA) provides an excellent account of the dynamic
spin correlation function for a dominant triplon. The starting point
is the following low temperature approximation for ${\cal S({\bf
Q},\omega})$, which concentrates all spectral weight in a resonant
mode.
\begin{equation}
{\cal S}^{\alpha\beta}({\bf Q},\omega)={\cal S}({\bf Q})\delta
(\hbar\omega-\epsilon ({\bf Q}))\delta_{\alpha\beta}. \label{sab}
\end{equation}

The first moment sum-rule\cite{hohenberg74:128} links the equal time
correlation function, ${\cal S}({\bf Q})$,  to the dispersion
relation as follows
\begin{eqnarray}\label{SMAb}
    {\cal S}({\bf Q})&=& -{2 \over 3}{1 \over \epsilon ({\bf Q})}[
     J_{\bf d}\langle {\bf S}_{1,i}\cdot {\bf S}_{2,i}\rangle
     ( 1-\cos ({\bf Q}\cdot {\bf d})) ].
\end{eqnarray}
Here we make the simplifying approximation that the molecular bond
with displacement vector $\bf d$ dominates the structure factor.
Within this framework, we examine different models of magnetism in
Cu(Quinoxaline)Br$_2$ by varying parameters in the following
phenomenological dispersion relation that allows for inter-molecular
interactions along the $(\bf{a}\pm\bf{b})/2$ and $\bf{c}$
directions:
\begin{eqnarray}
\epsilon({\bf Q})= \epsilon_{\bf b}(k)&+&2B_{hk}\cos2\pi h\cdot\cos
2\pi
k \\
&+&B_l\cos(2\pi l)
\end{eqnarray}
Here ${\bf Q}=h{\bf a}^*+k{\bf b}^*+l{\bf c}^*$. For models that
approach the spin ladder limit we use the following strong rung
coupling perturbation expansion,\cite{Reigrotzki94:6}
\begin{eqnarray}\label{1ddis}
\epsilon_{\bf b}(k)&=& J_{\perp}
[1+\frac{J_{\parallel}}{J_{\perp}}\cos 2\pi k
+\frac{1}{4}(\frac{J_{\parallel}}{J_{\perp}})^2(3-\cos 4\pi k)\nonumber\\
&&-\frac{1}{8}(\frac{J_{\parallel}}{J_{\perp}})^3(2\cos
2\pi k+2\cos 4\pi k -\cos 6\pi k -3)\nonumber\\
&&+O(\frac{J_{\parallel}}{J_{\perp}})^4],
\end{eqnarray}
For models where the dispersion along $\bf b$ is not dominant we use
the simpler form
\begin{equation}
\epsilon_{\bf b}(k)=B_0+B_k\cos 2\pi k
\end{equation}
The spherical average of ${\cal S}({\bf Q},\omega)$
(Eq.~(\ref{powderavg})) was in general calculated numerically except
for truly one-dimensional models where we used the analytical
result\cite{Hammar97:57}.

\subsection{One-Triplon CUT Model}
For an ab-initio description of the dynamic spin correlation
function for the two-leg ladder, we use a particle conserving
Continuous Unitary Transformation
(CUT).\cite{knett00,knett01,knett03a,knett04} The CUT is realized in
a perturbative fashion about the limit of isolated rung dimers and
can yield the spectral weight and dispersion of the triplon
elementary excitations, as well as the multi-triplon contributions.
Here we are interested in the low-energy part of the dynamical
structure factor for $x=J_\parallel/J_\perp \leq 1$. In this regime
the dynamical structure factor is dominated by the one-triplon part.
Two-triplon contributions are located at higher energies and carry
only a small part of the total spectral weight ($10\%$ for $x=0.6$).
It is thus justified to restrict the discussion to the one-triplon
part of the dynamical structure factor, which is characterized by
the one-triplon dispersion $\epsilon_{\rm \small CUT}({\bf Q})$ and
the one-triplon spectral weight $a^2({\bf Q})$. The high-order
series for both quantities are expressed in terms of an internal
parameter of the system, the energy gap $\Delta$, instead of $x$,
and quantitative results for $x \leq 1$ are obtained by
extrapolation.\cite{schmi03a,schmi03d}

The dynamic spin correlation function at $T=0$ and for $\omega>0$ is
given by
\begin{eqnarray}
{\cal S}_{\rm \small CUT}(\tilde{q},\omega)=-\frac{1}{\pi}\Im m\frac{a^2(\tilde{q})}{\hbar\omega-\epsilon_{\rm \small CUT}(\tilde{q})+i\delta}\nonumber \\
=\frac{1}{\pi}\frac{a^2(\tilde{q})\delta}{(\hbar\omega-\epsilon_{
\rm \small CUT}(\tilde{q}))^2+\delta^2}.
\end{eqnarray}

Here $\delta=0.02$ meV was used  to avoid divergences without
introducing broadening beyond the experimental resolution. The
spherical average of the dynamic spin correlation function is
obtained as follows:
\begin{eqnarray}
\widetilde{\cal S}_{\rm {\tiny CUT}}(Q,\omega) &=&
\frac{1}{4\pi}\int_{0}^{\pi}\sin\theta d\theta \int_{0}^{2\pi}d\phi
\\ \nonumber &\times& 2S_{\rm
{\tiny CUT}}(Q_{\parallel}b,\omega)(1-\cos(dQ_{\perp}\sin\phi)).
\end{eqnarray}

The integral over $\phi$ can be computed analytically leading to
\begin{eqnarray}\label{cut}
\widetilde{{\cal S}}_{\rm {\footnotesize CUT}}(Q,\omega) &=&
\frac{2}{bQ}\int_{0}^{Qu}dy S_{\rm {\footnotesize CUT}}(y,\omega) \\
\nonumber &\times& [1-J_0(\frac{d}{b}\sqrt{(b^2Q^2-y^2)})],
\end{eqnarray}
where $y=Qb\cos\theta$ and $J_0(x)$ is the zeroth order Bessel
function. From the structure of the material, we have $d=3.75$ $\AA$
and $\rm b=6.93$ $\AA$.

\subsection{Comparison between theory and data}

We now compare each of these models of magnetism in $\rm
Cu(Quinoxaline)Br_2$ to the observed magnetic scattering intensity.
Multiple scattering involving inelastic magnetic scattering followed
or preceded by incoherent elastic nuclear scattering was added to
the calculated single event scattering as described by
Eq.~(\ref{multip}) with the experimentally determined effective
transmission ${\cal T}=0.4$. We also allowed for an overall additive
constant to account for any  discrepancies in the background
subtraction.

\begin{figure}
\includegraphics[width=8.5cm,bbllx=80,bblly=370,bburx=535,
  bbury=695,angle=0,clip=]{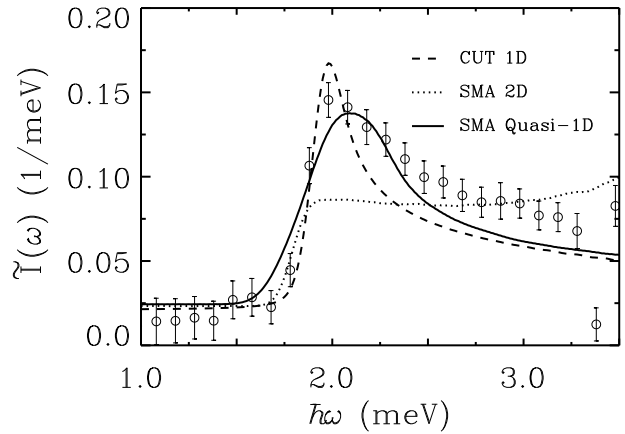}
\caption{Energy dependence of the normalized magnetic scattering
intensity for $\rm Cu(quinoxaline)Br_2$ averaged over wave vectors
0.5 $\rm \AA^{-1}\leq Q\leq$1.0 $\rm \AA^{-1}$. Open circles are
experimental data. The dashed line indicates the one-triplon dynamic
spin correlation function from the Continuous Unitary Transformation
theory. The dotted line shows the spectrum for a two dimensional
model with $\emph{B}_{0}=3.82$ meV and $\emph{B}_{hk}=1.02$ meV. The
solid line shows the spectrum for a quasi-one-dimensional ladder
model with $\emph{B}_{0}=3.82$ meV, $\emph{B}_{k}=1.84$ meV,
$\emph{B}_{hk}=0.08$ meV, and $\emph{B}_l=0.14$ meV. All models were
convolved with the experimental resolution function and an overall
constant was fit to account for any discrepancies in the background
subtraction procedure.} \label{ie}
\end{figure}

\subsubsection{Q and E cuts}
For an overview, Fig.~\ref{ie} shows the energy dependence of the
magnetic scattering averaged over wave-vectors from 0.5 $\rm
\AA^{-1}$ to 1.0 $\rm \AA^{-1}$ as follows:
\begin{equation}
\widetilde{I}(\omega)=\frac{\int Q^2\widetilde{I}(Q,\omega)dQ}{\int
Q^2dQ},
\end{equation}
These data are a measure of the magnetic density of states and are
sensitive to the dimensionality of the spin system. The dashed line
in Fig.~\ref{ie} shows a fit to the one dimensional CUT spectrum.
The peak in the data is broader than this truly 1D model. The dotted
line shows the spectrum for a two dimensional model where
$\emph{B}_0=3.82$ meV and $\emph{B}_{hk}=1.02$ meV, which produce a
gap in the spectrum that is consistent with the data. This 2D model
does not produce a peak at the spectral onset as observed in the
experiment. A quasi-one-dimensional model (solid line in
Fig.~\ref{ie}) with $\emph{B}_0=3.82$ meV, $\emph{B}_{k}=1.84$ meV,
$\emph{B}_{hk}=0.08$ meV and $\emph{B}_l=0.14$ meV on the other hand
yields an acceptable fit. From this we may conclude that $\rm
Cu(Quinoxaline)Br_2$ is magnetically quasi one-dimensional.

\begin{figure}
\includegraphics[width=8.5cm,bbllx=70,bblly=355,bburx=540,
  bbury=690,angle=0,clip=]{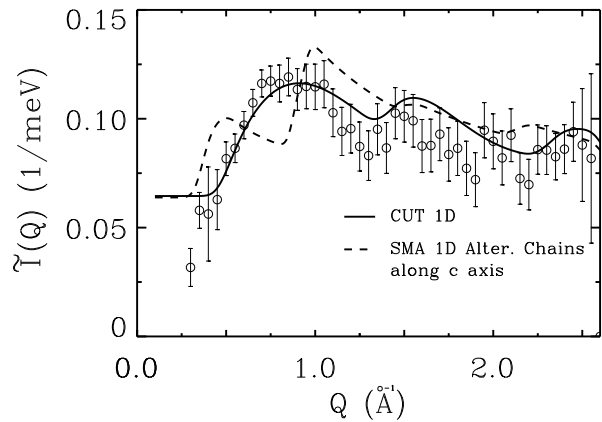}
\caption{Wave-vector dependence of normalized magnetic scattering
intensity for $\rm Cu(quinoxaline)Br_2$ at 1.4 K, integrated over
energy transfers 1.85 meV $\leq\hbar\omega\leq 2.65$ meV. Open
circles are experimental data.
 The solid and dash lines indicate the dynamic spin correlation
 function of the one-triplon CUT and SMA 1D alternating spin chain models, respectively,
 convolved with the experimental resolution function. Multiple scattering was
  added to the models as calculated from Eq.~(\ref{multip}) using the experimentally determined
  value of ${\cal T}=0.40$. Note that data for $Q<0.5$ $\AA^{-1}$ does not include the full energy range
  due to the kinematic limitations of the experiment (see Fig.~\ref{all}).} \label{iq}
\end{figure}

To establish the relevant one dimensional model, we examine the
wave-vector dependence of magnetic scattering averaged over the
energy range from 1.85 meV to 2.65 meV (Fig.~\ref{iq}).
$\widetilde{I} (Q)$ is sensitive to spatial spin correlations and in
particular the intra-dimer spin spacing, $d$. The data show a
rounded maximum at $Q_{0}\simeq0.8$ $\rm \AA^{-1}$, indicating that
the spins forming the dominant correlated spin pairs are separated
by $ d\approx\pi/Q_0=4$ $\rm \AA$. For comparison the nearest
neighbor Cu-Cu separation within the $\rm Cu_2Br_4$ molecular unit
is 3.75 $\rm \AA$. The solid line in Fig.~\ref{iq} shows the CUT
ladder model prediction. The excellent agreement with the data is
strong support for the ladder model. Conversely the $\bf{c}$-axis
alternating chain model for one-dimensional magnetism in $\rm
Cu(Quinoxaline)Br_2$ produces the dashed line which is clearly
inconsistent with the data.

\subsubsection{Intensity contour maps and global fits}
We performed global fits of the dynamic spin correlation function
associated with the models described in section $\rm IV$ to the $Q$
and $\omega$ dependent magnetic neutron scattering from $\rm
Cu(Quinoxaline)Br_2$. The results are summarized in Figs.~\ref{ie}
-~\ref{all}. A least-square fit of the ladder single-mode
approximation gave $\langle {\bf S}_{1,i}\cdot {\bf
S}_{2,i}\rangle=-0.37(1)$, $\langle {\bf S}_{1,i}\cdot {\bf
S}_{1,i+1}\rangle=-0.04(1)$. The former value is close to the
expectation value of -3/8 for a spin dimer singlet ground state. We
also found $J_{\parallel}=1.98(4)$ meV, $J_{\perp}=3.05(1)$ meV. The
ratio $J_{\parallel}/J_{\perp}\approx0.65$ indicates that the
perturbative expression for the dispersion relation in
Eq.~(\ref{1ddis}) is at the limit of applicability. Allowing for
inter-ladder coupling through finite values for $\emph{B}_{hk}$ and
$\emph{B}_l $ reduced the $\chi^2$ from 1.3 to 1.1 for
$\emph{B}_{hk}=0.08$ meV and $\emph{B}_{l}=0.14$ meV, which
indicates that there are weak interactions between ladders.
Fig.~\ref{all}(b) shows the best fit which corresponds to
$\emph{B}_0=3.82$ meV, $\emph{B}_k=1.84$ meV, $\emph{B}_{hk}=0.08$
meV, $\emph{B}_l=0.14$ meV, and $J_{\bf d}\langle {\bf S}_{1,i}\cdot
{\bf S}_{2,i}\rangle=-1.0$ meV.

\begin{figure}
\includegraphics[width=8.5cm,bbllx=100,bblly=375,bburx=525,
  bbury=690,angle=0,clip=]{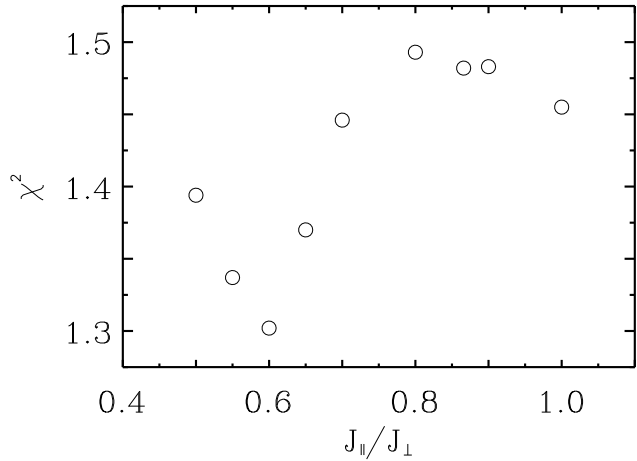}
\caption{$\chi^2={1\over N_{free}}\displaystyle
\sum_{j}(I_j^{obs}-I_j^{ex})^2/\sigma_{j}^2$ versus the ratio of
rail to rung exchange in the comparison of magnetic neutron
scattering data for \protect$\rm Cu(quinoxaline)Br_2$ to the
Continuous Unitary Transformation theory of one-triplon excitations
in a {\bf b}-axis spin ladder.} \label{chi2}
\end{figure}

As opposed to the strong coupling perturbation expansion, the CUT
can provide results for $\rm {\cal S}(Q,\omega)$ for
$J_{\perp}\approx J_{\parallel}$. Fits were undertaken for fixed
exchange ratios $x=J_{\parallel}/J_{\perp}$ ranging from 0.5 to 1.
For each value of $x$, the only adjustable parameters was an overall
scale factor  and the constant background. $J_{\perp}$ was selected
so as to produce the observed value of $\Delta$ for each value of
$x$ as listed in Table~\ref{table1}. Fig.~\ref{chi2} shows the $x$
dependent smallest possible value for the $\chi^2$ goodness of fit
criterium. The best fit was obtained for $x=0.6$. The corresponding
values of the exchanges constants are $J_{\parallel}=1.98$ meV and
$J_{\perp}=3.3$ meV. The entire powder averaged dynamic spin
correlation function for this model is shown in Fig.~\ref{whole}(a).
Fig.~\ref{all}(a) compares the CUT model calculations for $x=0.6$ to
the experimental data. Fig.~\ref{whole}(b) shows that the strong
coupling dispersion relation obtained from the SMA fits is in fact
also in good agreement with that derived from the CUT.

\begin{figure}
\includegraphics[width=8.5cm,bbllx=90,bblly=85,bburx=465,
  bbury=700,angle=0,clip=]{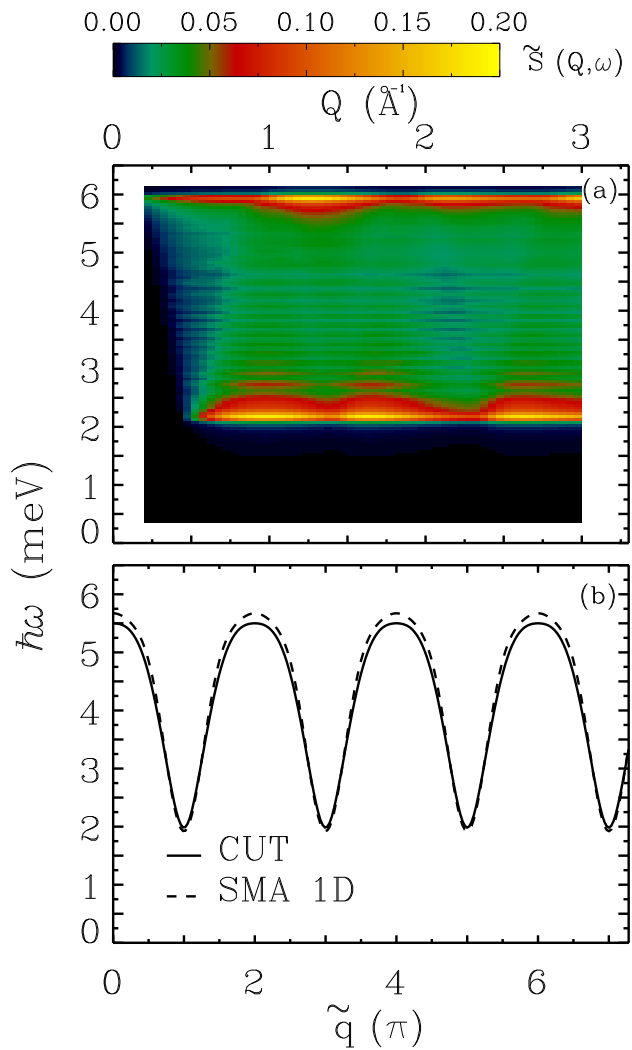}
\caption{(a) One-triplon contribution to the powder-averaged dynamic
spin correlation function for a spin ladder as calculated by the
Continuous Unitary Transformation. Lattice spacings are those
appropriate for $\rm Cu(quinoxaline)Br_2$ and the exchange constants
were derived by fitting to neutron scattering data as described in
the text. (b) The comparison of triplon dispersion relation for the
best fit between CUT (solid line) and SMA (dashed line) ladder
models.} \label{whole}
\end{figure}

\begin{table}
\caption{\label{tab:table1}Values of ladder exchange constants for
which the one triplon contribution to $\widetilde{S}(Q,\omega)$
was calculated for comparison to the data}
\begin{ruledtabular}
\begin{tabular}{lccr}
$J_{\parallel}$ (meV) & $J_{\perp}$ (meV)& $x=J_{\parallel}/J_{\perp}$& $\chi^2$ \\
\hline
1.56&3.12&0.5& 1.39 \\
1.77&3.21&0.55&1.34 \\
1.98&3.3&0.6 &1.30 \\
2.22&3.41&0.65 &1.37 \\
2.46&3.52&0.7 &1.45 \\
2.96&3.70&0.8 &1.49 \\
3.29&3.80&0.866 &1.48 \\
3.47&3.85&0.9 &1.48 \\
3.98&3.98&1.0 &1.46 \\
\end{tabular}
\end{ruledtabular}
\label{table1}
\end{table}

Whether the corresponding scattering cross section is approximated
by the CUT or the SMA, the ladder model clearly provides an
excellent account of the data. The virtue of CUT in this context is
that apart from the exchange constants and an overall scale factor
that in principle can be determined through the total moment sum
rule, there are no adjustable parameters.

\begin{figure}[t]
\includegraphics[width=8.5cm,bbllx=110,bblly=115,bburx=535,
  bbury=720,angle=0,clip=]{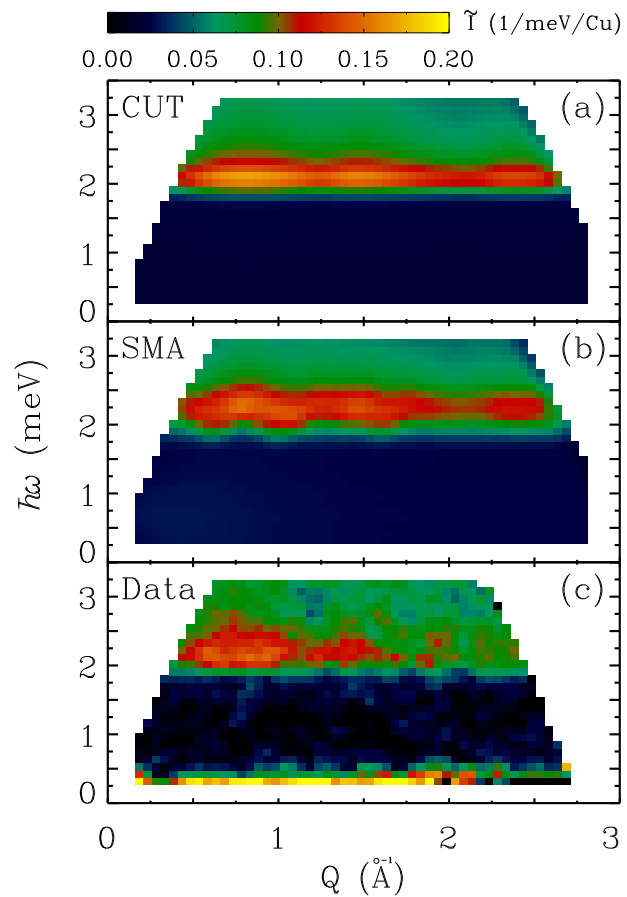}
\caption{Comparison of inelastic magnetic neutron scattering
$\widetilde{I}(Q,\omega)$ from $\rm Cu(Quinoxaline)Br_2$ at T=1.4 K
(c) to (a) the One-Triplon scattering derived from the Continuous
Unitary Transformation (b) a quasi-one-dimensional spin ladder model
in the Single-Mode Approximation. The models were complemented with
incoherent elastic double scattering, according to
Eq.~(\ref{phonon}) with ${\cal{T}}=0.40$ and an overall additive
constant was fit to account for any discrepancies in background
subtraction.} \label{all}
\end{figure}

\section{CONCLUSION}
In summary, we used neutron scattering to determine the origin of
quantum paramagnetism in $\rm Cu(Quinoxaline)Br_{2}$. The methods of
detailed balance correction, phonon subtraction, and multiple
scattering correction were described and used to extract the
relatively weak magnetic signal from the strong nuclear incoherent
scattering for quantitative comparison to theoretical models. The
observed spin gap $\Delta\approx1.9$ meV is consistent with magnetic
susceptibility and high-field magnetization
measurement.\cite{Landee03:22} The neutron scattering data are well
described  by the Continuous Unitary Transformation theory of a
spin-1/2 ladder along the {\bf b}-direction of $\rm
Cu(Quinoxaline)Br_{2}$ with exchange constants $J_{\parallel}=1.98$
meV and $J_{\perp}=3.3$ meV. There is also evidence of inter-ladder
dispersion with a bandwidth of approximately 0.2 meV. A {\bf c}-axis
alternating chain model was shown to be inconsistent with the data.
While there may exist other models that are consistent with the
present powder data, our results strongly support the interpretation
of $\rm Cu(Quinoxaline)Br_{2}$ as a quantum spin-1/2 ladder with
significant inter-rung coupling ($J_{\parallel}/J_{\perp}=0.6$).
Efforts are now under way to test this conclusion and explore
two-triplon dynamics in this interesting quantum paramagnet through
inelastic neutron scattering from an assembly of small crystalline
samples.

\begin{acknowledgments}
It is a pleasure to thank Rick Paul for help with neutron activation
analysis and Seunghun Lee for providing software to implement the
detailed balance correction. Work at JHU was supported by the NSF
through Grant No. DMR-0306940. This work utilized neutron research
facilities that were partially supported by NSF through DMR-0086210
and DMR-0454672.
\end{acknowledgments}



\end{document}